\documentclass[conference]{IEEEtran}
\usepackage[utf8]{inputenc}
\usepackage[T1]{fontenc} 
\usepackage{amsmath}
\usepackage[cmintegrals]{newtxmath}
\usepackage{bm} 
\usepackage{graphicx}
\usepackage{subcaption}
\usepackage{float}
\usepackage{multirow}

\title{Deep Learning of Cell Classification using Microscope Images of Intracellular Microtubule Networks}
\author{\IEEEauthorblockN{Aleksei Shpilman\IEEEauthorrefmark{1},
Dmitry Boikiy\IEEEauthorrefmark{1},
Marina Polyakova\IEEEauthorrefmark{2},
Daniel Kudenko\IEEEauthorrefmark{1}\IEEEauthorrefmark{3},
Anton Burakov\IEEEauthorrefmark{4} and
Elena Nadezhdina\IEEEauthorrefmark{5}}

\IEEEauthorblockA{\IEEEauthorrefmark{1}Saint Petersburg National Research Academic University of the Russian Academy of Sciences,
St Petersburg, Russia\\ 
Email: aleksei@shpilman.com}
\IEEEauthorblockA{\IEEEauthorrefmark{2}Moscow Institute of Physics and Technology, Dolgoprudny, Russia\\}
\IEEEauthorblockA{\IEEEauthorrefmark{3}Department of Computer Science, University of York, UK\\}
\IEEEauthorblockA{\IEEEauthorrefmark{4}A.N. Belozersky Intstitute of Physico-Chemical Biology, Lomonosov Moscow State University, Moscow, Russia\\}
\IEEEauthorblockA{\IEEEauthorrefmark{5}Institute of Protein Research of the Russian Academy of Sciences, Puschino, Russia\\}}

\date{June 2017}

\begin{document}

\maketitle

\begin{abstract}
    Microtubule networks (MTs) are a component of a cell that may indicate the presence of various chemical compounds and can be used to recognize properties such as treatment resistance. Therefore, the classification of MT images is of great relevance for cell diagnostics. Human experts find it particularly difficult to recognize the levels of chemical compound exposure of a cell. Improving the accuracy with automated techniques would have a significant impact on cell therapy. In this paper we present the application of Deep Learning to MT image classification and evaluate it on a large MT image dataset of animal cells with three degrees of exposure to a chemical agent. The results demonstrate that the learned deep network performs on par or better at the corresponding cell classification task than human experts. Specifically, we show that the task of recognizing different levels of chemical agent exposure can be handled significantly better by the neural network than by human experts.
\end{abstract}

\section{Introduction}

In recent years, machine learning has become an important tool for various tasks involving biological and medical data, such as analysis of genetic code \cite{bioinf}, automated diagnosis\cite{diagnosis}, and medical image analysis such as MRI and X-ray scans and ultrasound images \cite{imaging}, tissue samples \cite{hist} and even images of single cells \cite{yeast}. However, the analysis of intra-cellular components, such as the cytoskeleton, and microtubule networks in particular, is not yet widely covered by machine learning applications. This is partially due to the lack of sufficiently large datasets suitable for machine learning. Our paper addresses this research gap by collecting a large dataset of microtubule network images and demonstrating the successful application of deep learning to intra-cellular analysis.  

Identifying the responsiveness of cells to chemical agents forms an important part of cell therapy, for example targeted chemo-therapy for cancer \cite{cancer}. Chemical agents affecting microtubules form a significant part of the active ingredients of chemotherapy drugs, for example agents like Taxol vinblastine, (Paclitaxel), Vinblastine, Vincristine, etc. These agents are classified as mitostatics, i.e. they effectively block cancer cell division. The problem is that many microtubule agents negatively influence the nervous system. That is why we also screen for new agents which should influence microtubules but have minimum side effects. For that search we need to evaluate microtubule system state in cells that are affected by potential agents. Microtubule networks are cell components whose images have been shown in past research to be particularly useful for recognizing the impact of chemical agents on a cell \cite{radial, neurodegeneration}. Such visual recognition tasks are normally carried out by human experts, but at times this poses a difficult challenge for them. Human expert classification accuracy is particularly low when it comes to distinguishing different levels of exposure to a chemical agent rather than just its presence. 

Recent research on automating microtubule network image classification \cite{radial} attempted to use expert-defined heuristics, but the performance of the automated system on the classification tasks was strictly worse than that of human experts. In this paper, we demonstrate the successful use of deep neural network learning techniques for the microtubule image classification task. The main contributions of our work are as follows: 

\begin{itemize}
    \item We demonstrate that deep learning techniques perform on par or better than human experts at microtubule network image classification tasks. Specifically, the trained convolutional neural network performed well on recognizing levels of chemical agent exposure of a cell, a task on which human experts typically achieve only poor levels of accuracy. 
    \item We collected a large dataset containing high resolution images of more than 3000 cells with visualized microtubule networks. To our knowledge, this is the largest existing dataset of microtubule networks of the same cell type under different conditions. We intend to make this dataset freely available to the public in the near future.
      
\end{itemize}

In the next Section, we provide some background on cell classification and the importance of microtubule networks. Following that, we briefly introduce deep learning, specifically convolutional neural networks (CNNs) and applications to biological image analysis. We then present our chosen approach to deep learning, followed by a discussion of the dataset, the experimental setup, and the results. Overall conclusions are drawn in the last section. 

\section{Cell classification and microtubule networks}

A cytoskeleton is a complex network of interlinking filaments and tubules that extends through intracellular space. It has many functions, such as giving cells their shape, transducting intracellular signals and various compounds from cell center to its periphery and otherwise\cite{molbiol}. Patterns in the intracellular signaling are indicative of neurological diseases such as Alzheimers and ALS \cite{neurodegeneration}. Also, during cell division, the cytoskeleton is involved in evenly distributing genetic material between two daughter cells by way of pulling chromosomes apart \cite{molbiol}, and thus are important for the detection of cancer \cite{hist}.

\begin{figure}[htpb]
    \centering
    \includegraphics[width=0.45\textwidth]{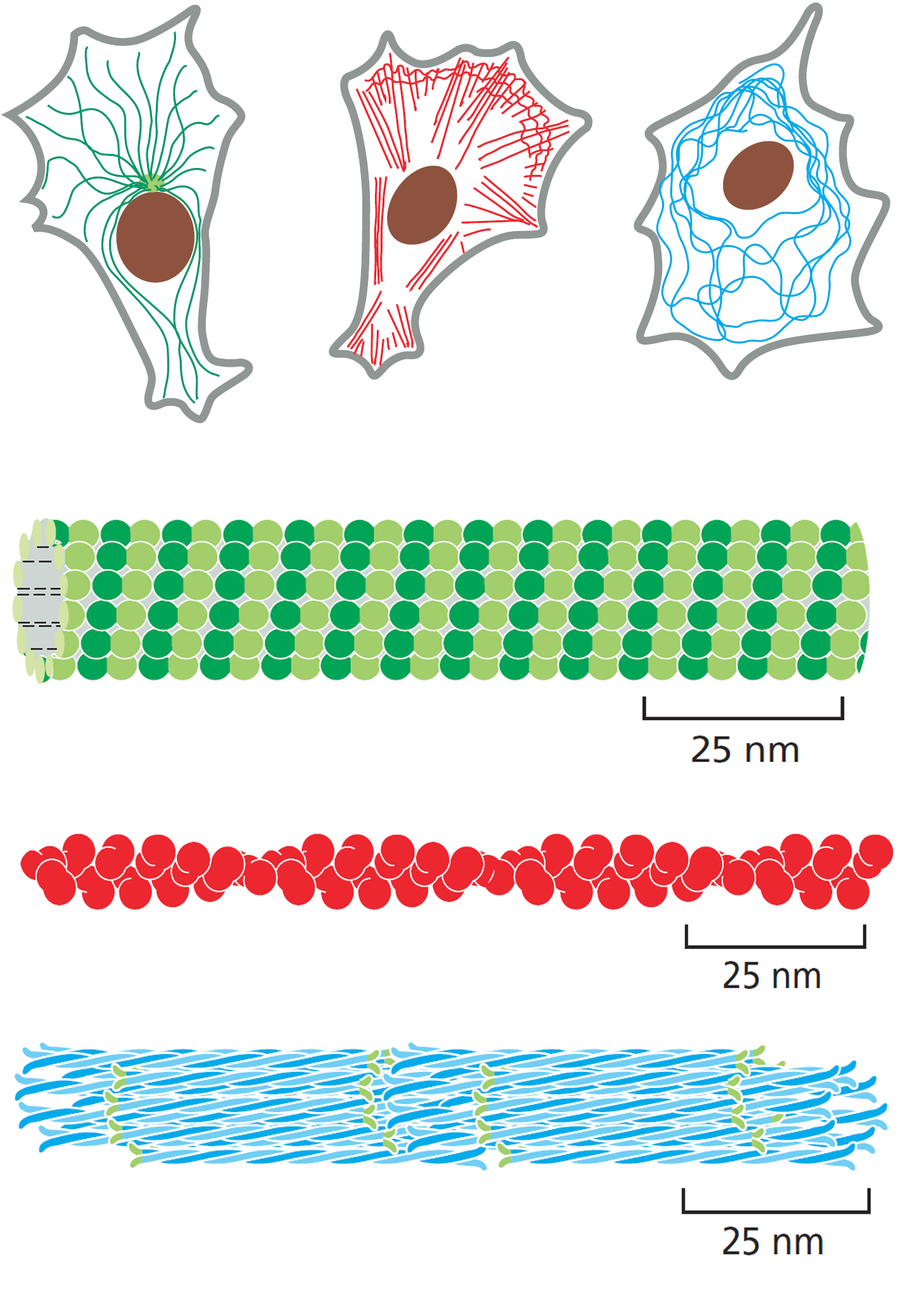}
    \caption{Cytoskeletal components and their structure. A  cytoskeleton consists three different types of biopolimers: microtubules (green), actin filaments (red) and intermidiate filaments (blue). In normal conditions microtubules form a radial structure. (From Alberts et al.\cite{molbiol})}
    \label{fig:micr}
\end{figure}

\begin{figure}[htpb]
    \centering
    \begin{subfigure}[b]{0.4\textwidth}
        \includegraphics[width=\textwidth]{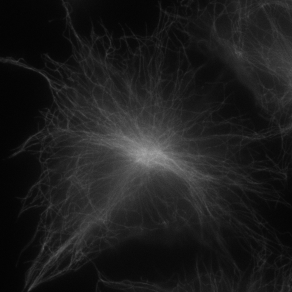}
        \caption{No agent}
        \label{fig:comp_c}
    \end{subfigure}
    ~ 
    \begin{subfigure}[b]{0.4\textwidth}
        \includegraphics[width=\textwidth]{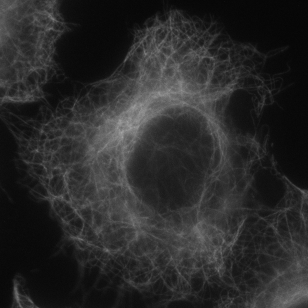}
        \caption{0.1M paclitaxel}
        \label{fig:comp_01}
    \end{subfigure}
    ~ 
    \begin{subfigure}[b]{0.4\textwidth}
        \includegraphics[width=\textwidth]{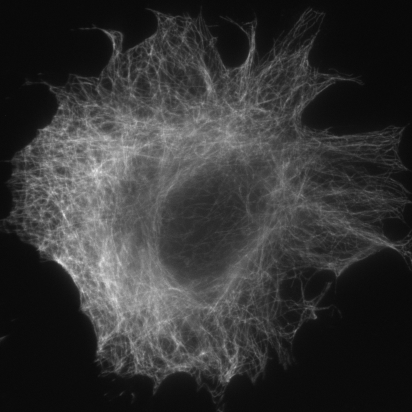}
        \caption{1M paclitaxel}
        \label{fig:comp_1}
    \end{subfigure}
    \caption{Microscopy images of cells with immunostained antibodies to tubulin for microtubule visualization.}
    \label{fig:comp}
\end{figure}

Microtubule networks are one of the primary components of the cytoskeleton, along with actin filaments and intermediate filaments (Figure \ref{fig:micr}). These are long, hollow cylinders made of the protein tubulin. In regular animal cells they have one end attached to a microtubule-organizing center thus forming a radial structure (as shown on Figure \ref{fig:comp_c}).

Microtubule networks are involved in the segregation of chromosomes during cell division, forming a mitotic spindle \cite{molbiol}. The tubulin family of proteins is a target for tubulin-binding chemo-therapeutics, which suppress the dynamics spindle to cause mitotic arrest and cell death, used in cancer treatment \cite{cancer}. Therefore, classifying microtubule networks in terms of reaction to different chemical agents will indicate how well the cancer cells will respond to chemo-therapy. Having this information along with genetic data, doctors can better decide which cancer treatment to employ and thus avoid ineffective chemo-therapy variants and their severe side-effects on the patient.

In order to find out the reaction of a cell to a chemical agent, images of the microtubule network are taken after the cell has been exposed to the chemical agent. If the cell reacted to the chemical agent, then the microtubule network will display certain kinds of disturbances in its geometry, for example tangled microtubules as shown in Figure \ref{fig:comp}.

Identifying such disturbances is at times difficult for human experts and often rather subjective. The main aim of the research presented in this paper is to develop an automated system that will improve the accuracy of microtubule network image classification as compared to human experts.  

\section{Deep learning for cell image analysis}

Deep Learning, and particulary Convolutional Neural Networks (CNN), currently forms the basis of most state-of-the-art image analysis with many significant successes, e.g. \cite{lenet, alexnet}. 

\begin{figure}[htpb]
    \centering
    \includegraphics[width=0.45\textwidth]{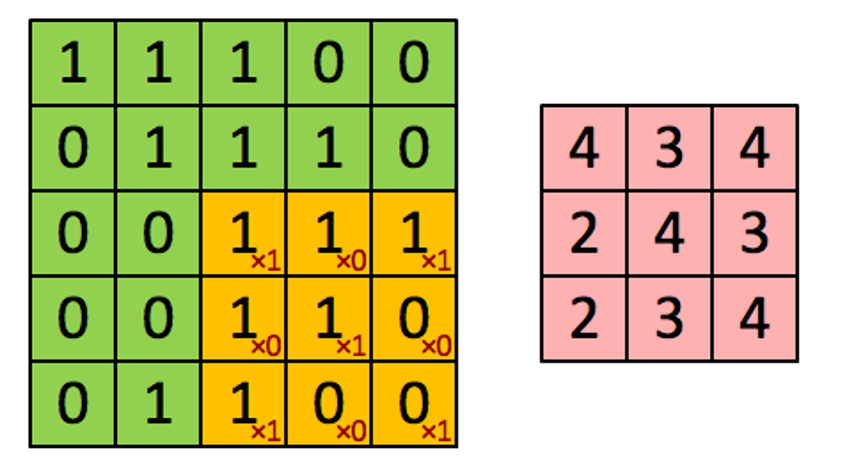}
    \caption{Convolution of input signal (left) to convolved features (right). At each pixel, convolution calculates the dot product between input (green) and kernel (red). In this case we apply 3x3 kernel with 1x1 stride.}
    \label{fig:convolution}
\end{figure}

CNNs \cite{lenet} were inspired by the organization of the animal visual cortex \cite{visual}. A CNN takes the image patch (convolution filter or kernel matrix) at each pixel as input and feeds it forward through a Neural Network by means of performing a dot product followed by a non-linear transformation. Convolutional layers compute the output of neurons that are connected to local regions in the input, each computing a dot product between their weights and a region they are connected to in the input (see Figure \ref{fig:convolution}).

The advantage of a CNN is that the features (weights of neural network connections) are automatically learned as kernel matrices, and therefore the importance of different visual clues is inferred directly from the classification task and not passed as heuristics from human experts. 

\begin{figure}[htpb]
    \centering
    \includegraphics[width=0.45\textwidth]{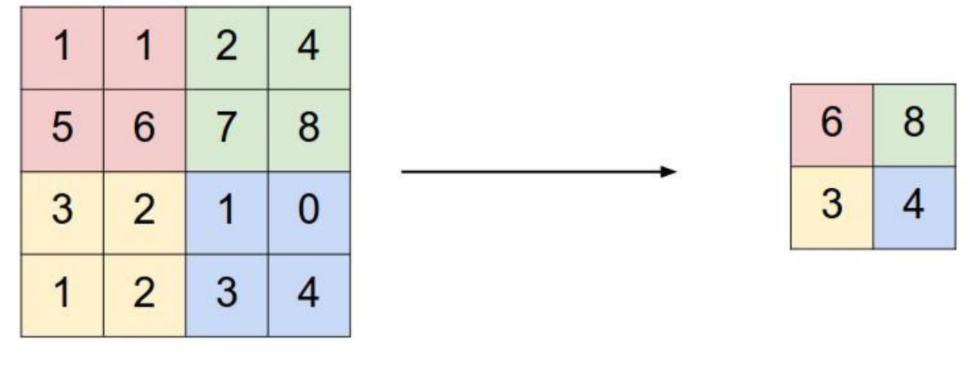}
    \caption{Max pooling of input signal (left) to max pool layer (right). At each step we take the maximum value from the filter of a given size. In this case we apply 2x2 filter with 2x2 stride.}
    \label{fig:maxpool}
\end{figure}

For dimensionality reduction and extraction of important features CNNs use maxpooling layers in which only the maximum output value is passed to the next layer (see Figure \ref{fig:maxpool}).

For the last layers, CNNs usually employ standard fully connected layers of neurons, as in classic Neural Networks.

In medical and biological image analysis, Deep Learning techniques have shown success in mitosis cell detection from histology images of breast cancer cells \cite{hist}. Chen et al. used CNNs for the automatic counting of immune cells on digitally scanned images of immunohistochemically stained cells and obtained correlation coefficients with the manual counts as high as 0.9949 \cite{imhc}. 

Apart from histology, Deep Learning has also been successfully applied to cell biology. For example, P{\" a}rnamaa and Parts \cite{yeast} achieved accurate classification of protein subcellular localization on microscopy images of yeast cells. 

So far, the application of Deep Learning to biological image analysis has been hampered by the lack of sufficiently large datasets. However, recent advances in imaging techniques which increase the speed of image production as well as achieving unprecedented levels of detail provide exciting opportunities of applying Deep Learning to cell biology. In this paper, we push this trend further by collecting the largest dataset (to our knowledge) of microtubule network images on a nano-meter scale and use it to train a CNN to classify them.

\section{Deep learning for Microtubule Network Classification}

For training the neural network we used Keras \cite{keras} with a Tensorflow backend \cite{tensorflow}. We experimented with a variety of arbitrary network topologies and parameter sets using cross-validation on the training data (excluding the test set). Specifically, we varied the following network topology parameters:

\begin{enumerate}
\item The number of convolutional layers, ranging from 1 to 5. 
\item Whether to use max pooling or not.
\item The number of filters in the convolutional layers, ranging from 8 to 256.
\item The size of the convolution kernel, ranging from 2x2 to 5x5.
\item Number of fully connected layers, ranging from 1 to 4.
\item Size of the fully connected layers, ranging from 16 to 512.
\end{enumerate}

The input image resolution was varied from 100x100 to 300x300 pixels. The number of neurons in the output layer depended on the classification task and was either 2 or 3. 

\begin{figure}[htpb]
    \centering
    \includegraphics[width=0.45\textwidth]{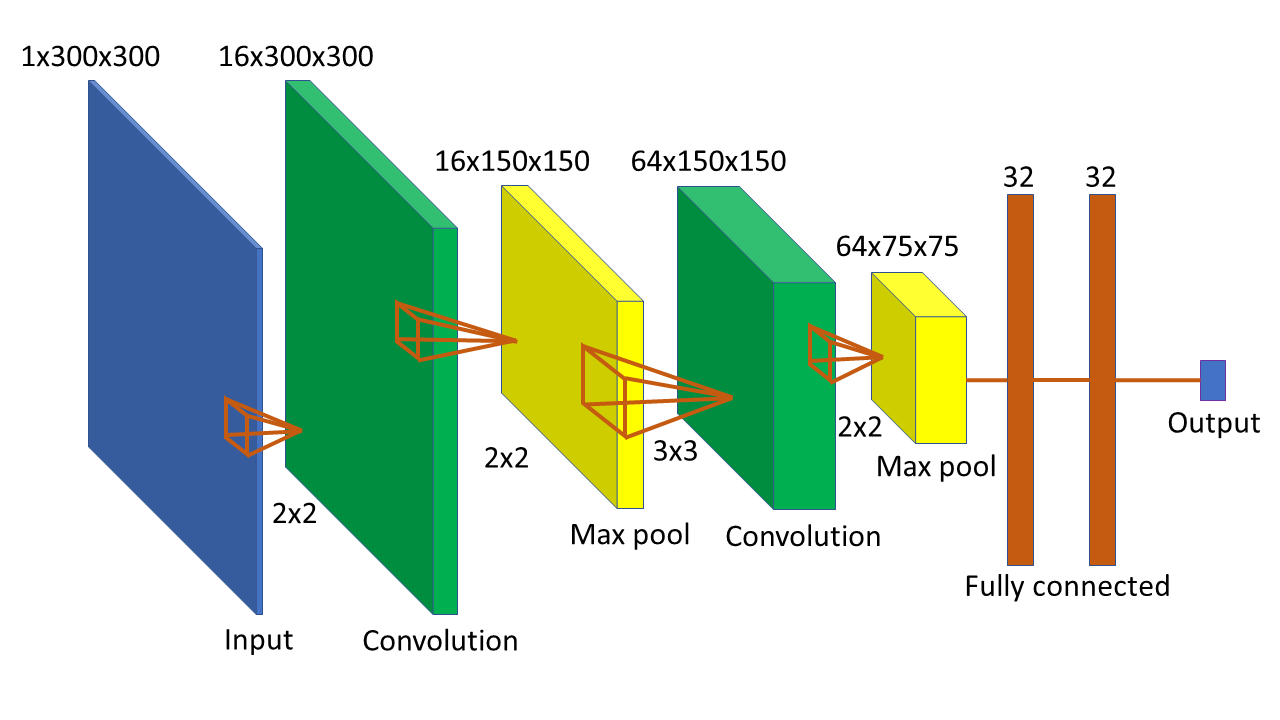}
    \caption{The best CNN topology as determined by cross-validation on the training data set.}
    \label{fig:topology}
\end{figure}

The resulting best topology choice is shown in Figure \ref{fig:topology}, and was as follows:

\begin{enumerate}
  \item Input layer --- grey scale 8bit image 300x300 pixels.
  \item Convolutional layer with 16 layers, 2x2 kernels and 1x1 stride.
  \item Max pooling layer with 2x2 kernel and 2x2 stride.
  \item Convolutional layer with 64 layers, 3x3 kernels and 1x1 stride.
  \item Max pooling layer with 2x2 kernel and 2x2 stride.
  \item Fully connected layer with 32 neurons.
  \item Fully connected layer with 32 neurons.
  \item Output layer --- softmax decision function with categorical cross-entropy as loss function.
\end{enumerate}

We have tried various non-linear transformations between layers with ReLU ($max(0,x)$) showing the best result. 

For regularization we used dropout and L2 weights regularization.

The best neural network parameters (as measured by cross validation on the training set) turned out to be as follows:

\begin{itemize}
    \item Optimizer --- "NAdam" ("Adam" \cite{adam} with Nesterov momentum).
    \item Learning rate --- 0.002.
    \item L2 regularization term --- 0.01.
    \item Dropout rate --- 0.5.
    \item Batch size --- 32.
\end{itemize}

\begin{figure}[htpb]
    \centering
    \begin{subfigure}[b]{0.22\textwidth}
        \includegraphics[width=\textwidth]{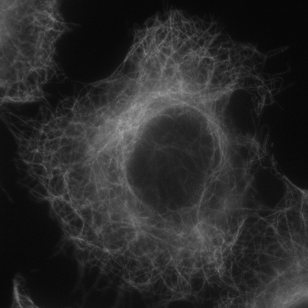}
        \caption{Original image}
        \label{fig:sharp_no_sharp}
    \end{subfigure}
    ~ 
    \begin{subfigure}[b]{0.22\textwidth}
        \includegraphics[width=\textwidth]{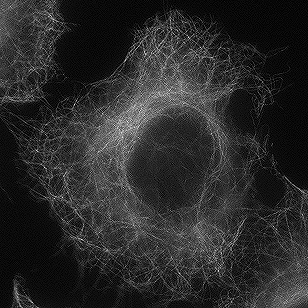}
        \caption{Sharpened image}
        \label{fig:sharp_sharp}
    \end{subfigure}
    \caption{Enhancing images with sharpening mask.}
    \label{fig:sharp}
\end{figure}

While this baseline network already performed comparatively well, we were able to further improve our approach by enhancing the data in the following ways:

\begin{enumerate}
  \item We applied three clockwise rotations to the images corresponding to 90, 180, and 270 degrees. Images that are rotated versions of each other were put either all in the training or all in the validation dataset.
  In this way we effectively quadrupled the size of our training data. Note that there were no rotated images in the test set. 
  \item We applied an image sharpening technique from the python library Pillow to the data to accent the linear structures (microtubules). The result of applying the sharpening mask is shown on Figure \ref{fig:sharp}. Note that while the sharpening technique helps the CNN, it does not help a human expert because the linear structures are already visible to the human eye in the original images. 
\end{enumerate}

\section{Experiments and Results}

In this section we provide more details on the datasets we collected, the experiment design, and the empirical results.

\subsection{Dataset}
In order to demonstrate the efficiency of our approach, we used the CV-1 cell line, derived from Cercopithecus aethiops monkey kidney cells. These cells are well spread and optimally suitable for microscopy which makes them particularly useful to obtain a large dataset. Also the basic properties of these cells are the same as those of human and other mammalian cells.

To visualize microtubules, the cells were fixed and immunostained with antibodies to tubulin --- the main component of microtubules. Specifically, as a primary antibody we used mouse monoclonal DM1A (Sigma-Aldrich, St. Louis, MO) and subsequently secondary antibodies conjugated to Fluorescein isothiocyanate, FITC (Jackson ImmunoResearch Laboratories; Newmarket, United Kingdom).

To implement changes in architecture of microtubules, we used paclitaxel --- chemotherapy medication agent which stabilizes microtubules, disrupts their dynamics and at high concentrations leads to the coalescence of microtubules into bundles. We used two different concentrations (0.1M and 1M) to see if independent experts and/or CNNs can detect not only the absence and presence of the chemical agent but also distinguish different reaction levels.

Following common biology experimentation practice, cells were separated into three groups for different treatment. The first group received no paclitaxel treatment (class "0"). The second and third groups received paclitaxel treatment in 0.1M (class "0.1") and 1M (class "1") concentrations. 

After the treatment, cells were fixed and immunostained, then photographed using an inverted fluorescent microscope Olympus IX71 with 63x PlanApo lens, supplied by 14-bit Olympus XM10 CCD-camera. Images were acquired with Micromanager software.

For each class we obtained 1000-1200 images of individual cells with a visualized microtubule network. As far as we are aware, this is the largest datasets of microtubule networks of the same cell type under different conditions ever collected. 

\subsection{Experiment Setup}

\subsubsection{Human Expert Experiments}

We separated 100 images from each class as a test set to evaluate the performance of our algorithm and compare it to that of human experts. For the latter comparison, we asked three experts from one of the leading laboratories in the field of microtubule studies to carry out a number of classification tasks. For each cell image in the test set, we specifically asked the experts to:

\begin{enumerate}
  \item Identify whether the cell is of class 0, 0.1, or 1.
  \item Identify whether the cell is of class 0 or 0.1. In case that the cell was really of class 1, we ignored the experts' answer. 
  \item Identify whether the cell is of class 0 or 1. In case that the cell was really of class 0.1, we ignored the experts' answer. 
  \item Identify whether the cell is of class 0.1 or 1. In case that the cell was really of class 0, we ignored the experts' answer. 
\end{enumerate}

Table \ref{tab:experts} shows the average accuracy of the the experts, accuracy of the highest scoring expert and accuracy of a majority voting method where each experts opinion is given one vote. As can be seen from the results, most experts were quite good at distinguishing cells with no chemical agent from a cell with any concentration of paclitaxel. A deeper analysis of the results showed that most problems the experts had with classification arose from fragmented and out of focus images, something that can't be fully avoided with current imaging technology. However, distinguishing between different concentrations of the effecting chemical agent proved to be an almost impossible task for a human.

\begin{table}[!ht]
\renewcommand{\arraystretch}{1.3}
\caption{Experts classification accuracy.}
\label{tab:experts}
\centering
\begin{tabular}{|c||c|c|c|}
\hline
Dataset & Average & Best expert & Voting\\
\hline
\hline
3 classes & 60.88 & 62.67 & 62.00\\
\hline
0 vs 0.1 & 88.17 & 90.50 & 89.00\\
\hline
0 vs 1 & 89.34 & 93.00 & 90.50\\
\hline
0.1 vs 1 & 50.17 & 52.00 & 51.00\\
\hline
\end{tabular}
\end{table}

As can be seen from the confusion matrices in Tables \ref{tab:experts_conf3} and \ref{tab:experts_conf}, the main difficulty for experts was differentiating between classes "0.1" and "1". We believe that this may be due to the fact that not all cells display a drastic visual difference between lower and higher concentrations of the chemical agent. In the case of small increases in network entanglement for a class "1" cell, these are not detected by the human eye and experts tend to classify such images as class "0.1". 

\begin{table}[!ht]
\renewcommand{\arraystretch}{1.3}
\caption{Confusion matrix for voting experts for the 3 class recognition task.}
\label{tab:experts_conf3}
\centering
\begin{tabular}{l|l|c|c|c}
\multicolumn{2}{c}{} & \multicolumn{3}{c}{True class}\\
\cline{3-5}
\multicolumn{2}{c|}{} & 0 & 0.1 & \multicolumn{1}{c|}{1}\\
\cline{2-5}
\multirow{3}{*}{Expert guess} & 0 & 84 & 6 & \multicolumn{1}{c|}{3}\\
\cline{2-5}
& 0.1 & 16 & 58 & \multicolumn{1}{c|}{53}\\
\cline{2-5}
& 1 & 0 & 36 & \multicolumn{1}{c|}{44}\\
\cline{2-5}
\end{tabular}
\end{table}

\begin{table}[!ht]
\renewcommand{\arraystretch}{1.3}
\caption{Confusion matrices for voting experts for the 2 class recognition tasks.}
\label{tab:experts_conf}
\centering

\begin{tabular}{l|l|c|c}
\multicolumn{2}{c}{} & \multicolumn{2}{c}{True class}\\
\cline{3-4}
\multicolumn{2}{c|}{} & 0 &  \multicolumn{1}{c|}{0.1}\\
\cline{2-4}
\multirow{2}{*}{Expert guess} & 0 & 84 & \multicolumn{1}{c|}{6}\\
\cline{2-4}
& 0.1 & 16 & \multicolumn{1}{c|}{94}\\
\cline{2-4}
\end{tabular}

\vspace{5mm}

\begin{tabular}{l|l|c|c}
\multicolumn{2}{c}{} & \multicolumn{2}{c}{True class}\\
\cline{3-4}
\multicolumn{2}{c|}{} & 0 &  \multicolumn{1}{c|}{1}\\
\cline{2-4}
\multirow{2}{*}{Expert guess} & 0 & 84 & \multicolumn{1}{c|}{3}\\
\cline{2-4}
& 1 & 16 & \multicolumn{1}{c|}{97}\\
\cline{2-4}
\end{tabular}

\vspace{5mm}

\begin{tabular}{l|l|c|c}
\multicolumn{2}{c}{} & \multicolumn{2}{c}{True class}\\
\cline{3-4}
\multicolumn{2}{c|}{} & 0.1 &  \multicolumn{1}{c|}{1}\\
\cline{2-4}
\multirow{2}{*}{Expert guess} & 0.1 & 75 & \multicolumn{1}{c|}{70}\\
\cline{2-4}
& 1 & 25 & \multicolumn{1}{c|}{30}\\
\cline{2-4}
\end{tabular}
\end{table}

\subsubsection{Convolutional Neural Network Experiments}

For training the neural network we used 1000 sharpened images (4000 after the application of image rotations) from each class. Ten-fold cross-validation accuracy on the training set was used as the stopping criterion for training. If no improvement of accuracy was seen after two epochs of training, the process was stopped. 

We trained our network on an AWS p2.xlarge instance with an Nvidia K80 Tesla GPU with 12GB of GPU memory.

Table \ref{tab:neural} shows the performance of the CNN with training on the original dataset, the dataset enhanced with image rotations (T), and the dataset with rotations and sharpening (T+S). The results clearly show the usefulness of the dataset enhancements. Also, the table compares the CNN performance to that of the best human expert and human experts' majority voting. In the three-class categorization task, the CNN performs comparable to the best human expert and the human voting scheme, as measured by t-test. In the task of distinguishing different concentrations of the chemical agents, CNNs perform significantly better, improving classification accuracy from 52\% for the best performing human expert to 70.5\% (p=0.0001 on t-test). In the other two classification tasks detecting the mere presence of the chemical agent, namely "0" vs "0.1" and "0" vs "1", the CNN performs on par with the best human expert, mis-classifying only 4 more test examples (out of 200) than the human. In addition, we performed experiments with XGBoost \cite{xgboost}, a state-of-the-art tree-based ML package on the "3 classes" and the "0.1" vs "1" tasks and the accuracies were significantly lower, which shows the benefit of Deep Learning.  

\begin{table}[!ht]
\renewcommand{\arraystretch}{1.3}
\caption{CNN classification accuracy compared to human performance. T stands for the addition of image rotations and S for image sharpening.}
\label{tab:neural}
\centering
\begin{tabular}{|c||c|c|c|c|c|}
\hline
Dataset & Best Expert & Voting & CNN & CNN+T & CNN+T+S\\
\hline
\hline
3 classes & 62.67 & 62.00 & 57.67 & 59.00 & \textbf{66.00}\\
\hline
0 vs 0.1 & \textbf{90.50} & 89.0 & 79.50 & 84.00 & 88.50\\
\hline
0 vs 1 & \textbf{93.00} & 90.50 & 88.00 & 90.50 & 91.00\\
\hline
0.1 vs 1 & 52.00 & 51.00 & 55.00 & 56.00 & \textbf{70.50}\\
\hline
\end{tabular}
\end{table}

The confusion matrices in Tables \ref{tab:neural_conf3} and \ref{tab:neural_conf} clearly show that CNNs do not have the same difficulty as humans do in distinguishing different concentrations of the chemical agent.

\begin{table}[H]
\renewcommand{\arraystretch}{1.3}
\caption{Confusion matrix for CNN+T+S for 3 classes.}
\label{tab:neural_conf3}
\centering
\begin{tabular}{l|l|c|c|c}
\multicolumn{2}{c}{} & \multicolumn{3}{c}{True class}\\
\cline{3-5}
\multicolumn{2}{c|}{} & 0 & 0.1 & \multicolumn{1}{c|}{1}\\
\cline{2-5}
\multirow{3}{*}{CNN guess} & 0 & 87 & 27 & \multicolumn{1}{c|}{19}\\
\cline{2-5}
& 0.1 & 10 & 58 & \multicolumn{1}{c|}{28}\\
\cline{2-5}
& 1 & 3 & 15 & \multicolumn{1}{c|}{53}\\
\cline{2-5}
\end{tabular}
\end{table}

\begin{table}[H]
\renewcommand{\arraystretch}{1.3}
\caption{Confusion matrices for CNN+T+S.}
\label{tab:neural_conf}
\centering

\begin{tabular}{l|l|c|c}
\multicolumn{2}{c}{} & \multicolumn{2}{c}{True class}\\
\cline{3-4}
\multicolumn{2}{c|}{} & 0 &  \multicolumn{1}{c|}{0.1}\\
\cline{2-4}
\multirow{2}{*}{CNN guess} & 0 & 91 & \multicolumn{1}{c|}{14}\\
\cline{2-4}
& 0.1 & 9 & \multicolumn{1}{c|}{86}\\
\cline{2-4}
\end{tabular}

\vspace{5mm}

\begin{tabular}{l|l|c|c}
\multicolumn{2}{c}{} & \multicolumn{2}{c}{True class}\\
\cline{3-4}
\multicolumn{2}{c|}{} & 0 &  \multicolumn{1}{c|}{1}\\
\cline{2-4}
\multirow{2}{*}{CNN guess} & 0 & 90 & \multicolumn{1}{c|}{8}\\
\cline{2-4}
& 1 & 10 & \multicolumn{1}{c|}{92}\\
\cline{2-4}
\end{tabular}

\vspace{5mm}

\begin{tabular}{l|l|c|c}
\multicolumn{2}{c}{} & \multicolumn{2}{c}{True class}\\
\cline{3-4}
\multicolumn{2}{c|}{} & 0.1 &  \multicolumn{1}{c|}{1}\\
\cline{2-4}
\multirow{2}{*}{CNN guess} & 0.1 & 71 & \multicolumn{1}{c|}{30}\\
\cline{2-4}
& 1 & 29 & \multicolumn{1}{c|}{70}\\
\cline{2-4}
\end{tabular}
\end{table}

\section{Conclusions}

In this paper, we showed that Convolutional Neural Networks can perform significantly better than human experts on cell image classification tasks, which are important for cell therapy. This was demonstrated on a microtubule network image classification task, for which we collected a large dataset which was subsequently enhanced with rotations and sharpening techniques. 

In future work we intend to study the performance of CNNs on related cell classification tasks using a wider range of cell types and chemical agents. 

We believe that Deep Learning can help provide information about the best treatment regime for a patient from data such as microscopy images. It may be also useful to add patient information to the deep learning algorithm, such as genetic sequences, biochemical pathways, cell and tissue pathology, MRI and CAT scans and even data about patient's environment to create an ultimate personalized medicine approach. We hope that our work will serve as an inspiration to other researchers to collect and analyze more large biological image datasets to support this effort. 

\bibliography{bibliography}

\begin{thebibliography}{10}
\providecommand{\url}[1]{#1}
\csname url@samestyle\endcsname
\providecommand{\newblock}{\relax}
\providecommand{\bibinfo}[2]{#2}
\providecommand{\BIBentrySTDinterwordspacing}{\spaceskip=0pt\relax}
\providecommand{\BIBentryALTinterwordstretchfactor}{4}
\providecommand{\BIBentryALTinterwordspacing}{\spaceskip=\fontdimen2\font plus
\BIBentryALTinterwordstretchfactor\fontdimen3\font minus
  \fontdimen4\font\relax}
\providecommand{\BIBforeignlanguage}[2]{{%
\expandafter\ifx\csname l@#1\endcsname\relax
\typeout{** WARNING: IEEEtran.bst: No hyphenation pattern has been}%
\typeout{** loaded for the language `#1'. Using the pattern for}%
\typeout{** the default language instead.}%
\else
\language=\csname l@#1\endcsname
\fi
#2}}
\providecommand{\BIBdecl}{\relax}
\BIBdecl

\bibitem{bioinf}
I.~Inza, B.~Calvo, R.~Armananzas, E.~Bengoetxea, P.~Larranaga, and J.~A.
  Lozano, ``{{M}achine learning: an indispensable tool in bioinformatics},''
  \emph{Methods Mol. Biol.}, vol. 593, pp. 25--48, 2010.

\bibitem{diagnosis}
Z.~Obermeyer and E.~J. Emanuel, ``Predicting the future—big data, machine
  learning, and clinical medicine,'' \emph{The New England journal of
  medicine}, vol. 375, no.~13, p. 1216, 2016.

\bibitem{imaging}
B.~J. Erickson, P.~Korfiatis, Z.~Akkus, and T.~L. Kline, ``Machine learning for
  medical imaging,'' \emph{RadioGraphics}, vol.~37, no.~2, pp. 505--515, 2017.

\bibitem{hist}
D.~C. Cire{\c{s}}an, A.~Giusti, L.~M. Gambardella, and J.~Schmidhuber,
  \emph{Mitosis Detection in Breast Cancer Histology Images with Deep Neural
  Networks}.\hskip 1em plus 0.5em minus 0.4em\relax Berlin, Heidelberg:
  Springer Berlin Heidelberg, 2013, pp. 411--418.

\bibitem{yeast}
T.~P{\"a}rnamaa and L.~Parts, ``Accurate classification of protein subcellular
  localization from high-throughput microscopy images using deep learning,''
  \emph{G3: Genes, Genomes, Genetics}, vol.~7, no.~5, pp. 1385--1392, 2017.

\bibitem{cancer}
A.~L. Parker, M.~Kavallaris, and J.~A. McCarroll, ``Microtubules and their role
  in cellular stress in cancer,'' \emph{Frontiers in Oncology}, vol.~4, p. 153,
  2014.

\bibitem{radial}
E.~V. Usova, A.~V. Burakov, A.~A. Shpilman, and E.~S. Nadezhdina, ``Disturbance
  of the radial system of interphase microtubules in the presence of excess
  serum in cell culture medium,'' \emph{Biophysics}, vol.~53, no.~6, pp.
  523--526, Dec 2008.

\bibitem{neurodegeneration}
N.~J. Cairns, V.~M.-Y. Lee, and J.~Q. Trojanowski, ``The cytoskeleton in
  neurodegenerative diseases,'' \emph{The Journal of Pathology}, vol. 204,
  no.~4, pp. 438--449, 2004.

\bibitem{molbiol}
B.~Alberts, A.~Johnson, J.~Lewis, D.~Morgan, M.~Raff, K.~Roberts, and
  P.~Walter, \emph{Molecular Biology of the Cell, Sixth Edition:}.\hskip 1em
  plus 0.5em minus 0.4em\relax Taylor \& Francis Group, 2014.

\bibitem{lenet}
Y.~Lecun, L.~Bottou, Y.~Bengio, and P.~Haffner, ``Gradient-based learning
  applied to document recognition,'' \emph{Proceedings of the IEEE}, vol.~86,
  no.~11, pp. 2278--2324, Nov 1998.

\bibitem{alexnet}
A.~Krizhevsky, I.~Sutskever, and G.~E. Hinton, ``Imagenet classification with
  deep convolutional neural networks,'' in \emph{Advances in Neural Information
  Processing Systems 25}, F.~Pereira, C.~J.~C. Burges, L.~Bottou, and K.~Q.
  Weinberger, Eds.\hskip 1em plus 0.5em minus 0.4em\relax Curran Associates,
  Inc., 2012, pp. 1097--1105.

\bibitem{visual}
D.~H. Hubel and T.~N. Wiesel, ``Receptive fields and functional architecture of
  monkey striate cortex,'' \emph{The Journal of physiology}, vol. 195, no.~1,
  pp. 215--243, 1968.

\bibitem{imhc}
T.~Chen and C.~Chefd'hotel, \emph{Deep Learning Based Automatic Immune Cell
  Detection for Immunohistochemistry Images}.\hskip 1em plus 0.5em minus
  0.4em\relax Cham: Springer International Publishing, 2014, pp. 17--24.

\bibitem{keras}
F.~Chollet \emph{et~al.}, ``Keras,'' \url{https://github.com/fchollet/keras},
  2015.

\bibitem{tensorflow}
\BIBentryALTinterwordspacing
M.~Abadi, A.~Agarwal, P.~Barham, E.~Brevdo, Z.~Chen, C.~Citro, G.~S. Corrado,
  A.~Davis, J.~Dean, M.~Devin, S.~Ghemawat, I.~Goodfellow, A.~Harp, G.~Irving,
  M.~Isard, Y.~Jia, R.~Jozefowicz, L.~Kaiser, M.~Kudlur, J.~Levenberg,
  D.~Man\'{e}, R.~Monga, S.~Moore, D.~Murray, C.~Olah, M.~Schuster, J.~Shlens,
  B.~Steiner, I.~Sutskever, K.~Talwar, P.~Tucker, V.~Vanhoucke, V.~Vasudevan,
  F.~Vi\'{e}gas, O.~Vinyals, P.~Warden, M.~Wattenberg, M.~Wicke, Y.~Yu, and
  X.~Zheng, ``{TensorFlow}: Large-scale machine learning on heterogeneous
  systems,'' 2015, software available from tensorflow.org. [Online]. Available:
  \url{http://tensorflow.org/}
\BIBentrySTDinterwordspacing

\bibitem{adam}
D.~P. Kingma and J.~Ba, ``Adam: {A} method for stochastic optimization,''
  \emph{CoRR}, vol. abs/1412.6980, 2014.

\bibitem{xgboost}
\BIBentryALTinterwordspacing
T.~Chen and C.~Guestrin, ``Xgboost: {A} scalable tree boosting system,''
  \emph{CoRR}, vol. abs/1603.02754, 2016. [Online]. Available:
  \url{http://arxiv.org/abs/1603.02754}
\BIBentrySTDinterwordspacing

\end{thebibliography}
\bibliographystyle{IEEEtran}
\end{document}